\documentclass{article}

\usepackage[english]{babel}

\usepackage[a4paper,top=2cm,bottom=2cm,left=3cm,right=3cm,marginparwidth=1.75cm]{geometry}

\usepackage{amsmath}
\usepackage{graphicx}
\usepackage[colorlinks=true, allcolors=blue]{hyperref}

\title{SA-UNet for Retinal Vessel improvment using StyleGAN2}
\author{Nadav Potesman 201235629 nadavpotes@gmail.com \\ Ariel rechtman 206921983 rechtman1@mail.tau.ac.il}

\begin{document}
\maketitle

\begin{abstract}
Retinal vessels segmentation is well known problem in image processing on the medical field. Good segmentation may help doctors take better decisions while diagnose eyes disuses. This paper describes our work taking up the DRIVE challenge which include segmentation on retinal vessels. We invented a new method which combine using of StyleGAN2 and SA-Unet. Our innovation can help any small data set segmentation problem. 
\end{abstract}

\section{Introduction}

We introduce a method to improve the segmentation of retinal images by creating images and their corresponding segmentation maps. Our proposed solution include training a StylGAN2\cite{StyleGAN2} with the DRIVE dataset\footnote{https://drive.grand-challenge.org/Download/} and then segmenting the new synthesized images using SA-Unet\cite{SA-UNet} model which currently give State of the Art result on segmenting DRIVE images. We used the new generated images along with their segmentation in order to retrain SA-Unet to achieve better results.
\\
\\
Over the several years Neural Networks (NN) became the best way to deal with many computer vision problems.
While up to those days segmentation problems have been solved using Edge Detection and other standard image processing algorithms, NN have shown better performance. 
\\
\\
GANs networks have done a tremendous effect on image generators in the last few years. At 2019 Kerras et al from NVIDIA group had show some breakthrough development when they presented StyleGAN network\cite{DBLP:journals/corr/abs-1812-04948} which is a style-base generator.
StyleGAN images are very realistic, high-resolution and can be control easily throw latnet vectors.
\\ \\
Changlu Guo et al from Budapest University proposed a solution for DRIVE challenge.
They solution is one of the most felicitous solution to the challenge this days.
In this paper, the researchers propose a change of architecture of the U-Net\cite{DBLP:journals/corr/RonnebergerFB15} and rename the new model as SA-Unet.
The main difference between U-Net and SA-Unet is in the backbone of convolutions layers, and in the spatial attention block between the encoder and the decoder as shown in figure 1.
Those changes increase the performance of the SA-Unet especially in the tiny vessel segmentation.
The performance of SA-Unet on the DRIVE dataset is shown below (Table 1). \\

\begin{table}
\centering
\begin{tabular}{p{1.5cm}|p{1.5cm}|p{1.5cm}|p{1.5cm}|p{1cm}|p{1.5cm}}
Method & Sensativity & Specicity & Accuracy & AUC & F1 Score  \\\hline
SA-UNet &  0.8212  & 0.9840  & 0.9698 & 0.9864 & 0.8263 \\
U-Net & 0.7677 & 0.9857 & 0.9666 & 0.9789 & 0.8012 \\
IterNet & 0.7735 & 0.9838 & 0.9573 & 0.9816 & 0.8205
\end{tabular}
\caption{\label{tab:widgets}Performance comparison on the DRIVE dataset.}
\end{table}

\begin{figure}
\centering
\includegraphics[width=0.7\textwidth]{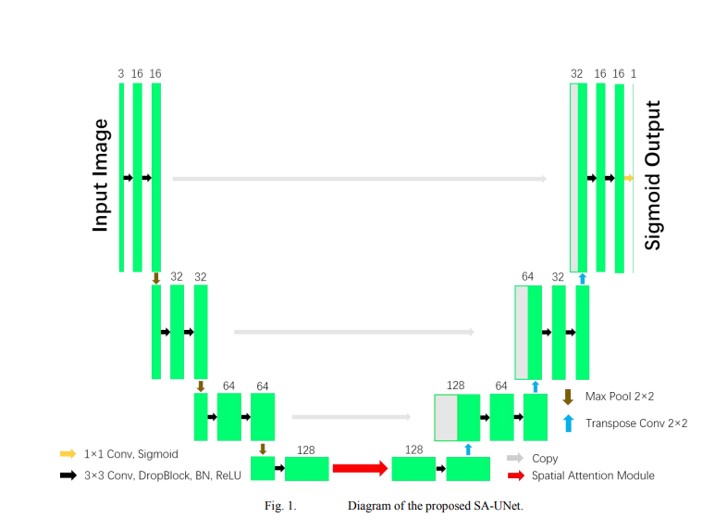}
\caption{\label{fig:SA-UNet}SA-UNet architecture.}
\end{figure}

\section{Related Work}
Small dataset have been always one of the most problematic fields for machine learning algorithms, and especially for Neural Network algorithms. The main familiar way to increase the size of the dataset is using augmentation such as crop, flip etc. \cite{DataAugmentation}.
Rential vessels blood are hard to segment, it always have been a big challenge, because the size of vessel may be from single to pixel up to ten pixels and more.
Even for normal human that mission consider to be a tough mission.\\ \\
U-Net \cite{DBLP:journals/corr/RonnebergerFB15} have done a great job in segment the data.
IterNet \cite{li2020iternet} network take advantage of U-Net by concatenate U-net with many mini-U-nets. IterNet action way is make at start rough segmentation map, and then improving the segmentation map by take apply the mini-U-nets on small regions.
\\ \\
SA-Unet \cite{SA-UNet} take the U-Net architecture but change the backbones of the convolutions layers and add the spatial attention between the encoder, and the decoder.\\ \\
Gal ofir et al. \cite{ofir2021improving} have done a step forward in that topic when they combine StyleGAN with Neural Network in propose to increase the size of the dataset, they took pre-trained IterNet and after generating synthetic images using StyleGan segment the images, after that they make manual corrections on the segmented data, and by that they have get over the problem of small dataset, with the price of less realistic images, and manual correction. \\

\section {Data}
The Digital Retinal Images for Vessel Extraction (DRIVE) dataset contains 20 images for train, and 20 images for test, images resolution is 565*584 in RGB format. Each Image have mask that cover the actual data of the Retinal. Train images contains a single manual segmentation of the vasculature. All human observers that manually segmented the vasculature were instructed and trained by an experienced ophthalmologist. They were asked to mark all pixels for which they were for at least 70 percent certain that were vessels.\\
The data also contains many various of eyes healthy ones and infected. We didn't make use of eyes status during our work. \\

\section {Methods}

\subsection{SA-Unet}
DRIVE dataset is well-known challenge which have been solved in many ways. SA-Unet is one of the SOTA solution. The big advantage of SA-Unet is that it combine U-Net, SD-Unet, and spacial attention block.
Despite it - they have a free well work code in github page\footnote{https://github.com/clguo/SA-UNet}, so it make the work more easily. The architecture of SA-Unet is very similar to U-Net. But SA-Unet have less parameters to optimize, so the training process is much shorter, and overfitting problem are less likely to happen.\\ \\
Dropout have been replace with dropblock which discard region of interest instead of discard randomly weights \cite{SA-UNet}, despite it, between any pair of convolutions layers there are Batch Normalization as shown in figure 2. Moreover SA-Unet came with spatial attention block between the encoder and the decoder. Attention have been in the last few years the more powerful technology in deep learning fields \cite{Attention4Segmentation_2019_CVPR}.\\ 

\subsection{StyleGAN}
Augmented data always have been a big computer vision problem. There are many ways to augmented data, start from more trivial ways such as image classical transforms such as flip, clip, rotate etc. those methods achieve nice performance at start, but there were still a need to create images that will be close to the real image, but the with major part of different.
Neural Networks open an opportunity to create a "deep fake" images, General Adversarial Networks(GAN) made an alternative way to create such images. Now days state of the art (SOTA) GAN is StyleGAN2 while in compare to SW-GAN and DCGAN \cite{radford2015unsupervised,arjovsky2017wasserstein}.

\subsection{Transfer Learning}
One of the best way to train a Neural Network model is take a pre-trained network, and train only the final layers. We examined transfer learning vs. train a completely new Network from scratch. The way we chose to evaluate the network images wad FID metric which given in eaquation 1.

\begin{equation}
    FID(x,g) = ||{(\mu_{x}-\mu_{g})^2}||_2 + Tr(\sigma_{x}+\sigma_{g} - 2(\sigma_{x}-\sigma_{g})^{1/2})
\end{equation}

FID metric measure the distance between x the origin image, and g the generated image, by transforming the images to a feature space, mu  stands for average, and sigma for covariance as in normal Guassian distribution, Tr stands for trace.\\
Our test had show that transfer learning is almost the same like training StyleGAN from scratch, the result of StyleGan synthesised image compare to real image can be show in figure 4.

\begin{figure}
\centering
\begin{minipage}{0.3\linewidth}
\includegraphics[width=\linewidth]{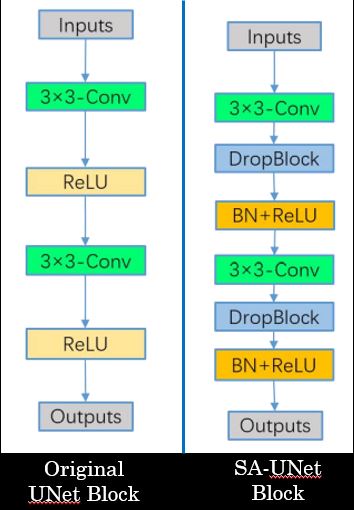}
\caption{\label{fig:DropBlock}Drop Block}
\end{minipage}
\hspace{.05\linewidth}
\begin{minipage}{0.55\linewidth}
\includegraphics[width=\linewidth]{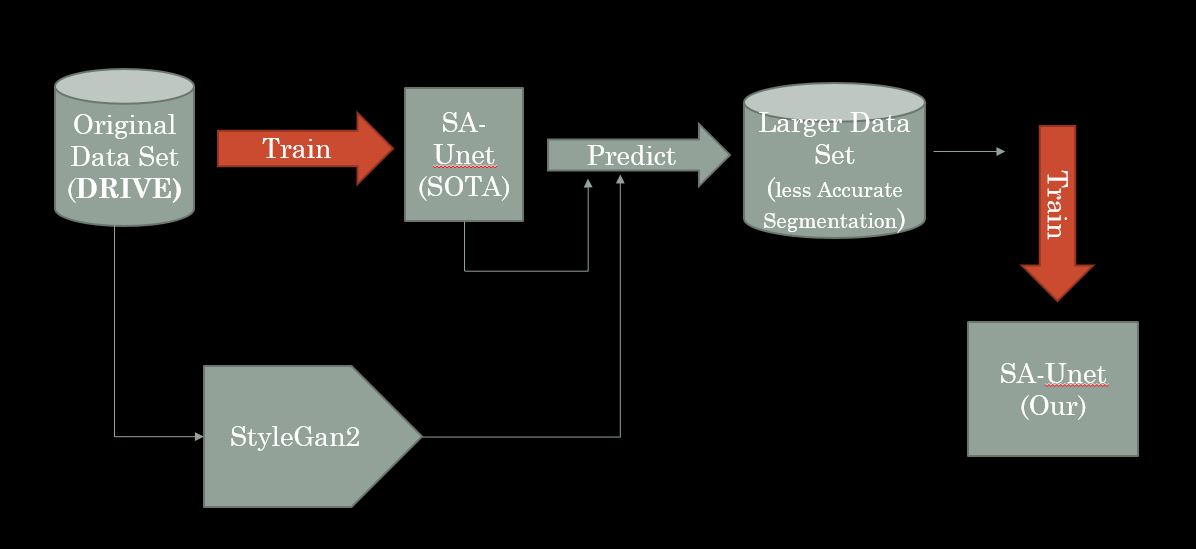}
\caption{\label{fig:Ensemble}Ensemble.}
\end{minipage}
\end{figure}

\subsection{Suggested Pipeline}
In our work we suggest an ensemble that may help all little dataset problems get better performance base on the same technique (figure 3).

\begin{enumerate}
\item Train the best Neural Network with the original dataset. 
\item Train StyleGAN2 with the original dataset. transfer learning are also a good option. 
\item Segment the synthetics generated image using the Net from part 1.
\item Train the same net from start using all the data.
\item (optional) Repeat section 2-4 until you get good enough results.
\end{enumerate}
One of our inference is that it recommended start train the SOTA net using the generated data, and do the final epochs using the original data. It make sense, because the original data is the most accurate segment so it will cause the best fine tuning.

\subsection {Ensamle}
Another approach we test on this paper is to use the above pipline together with the original SA-Unet in orer to create ensamle of networks. This will be described in detail in the next section. \\

\begin{figure}
\centering
\includegraphics[width=0.8\linewidth]{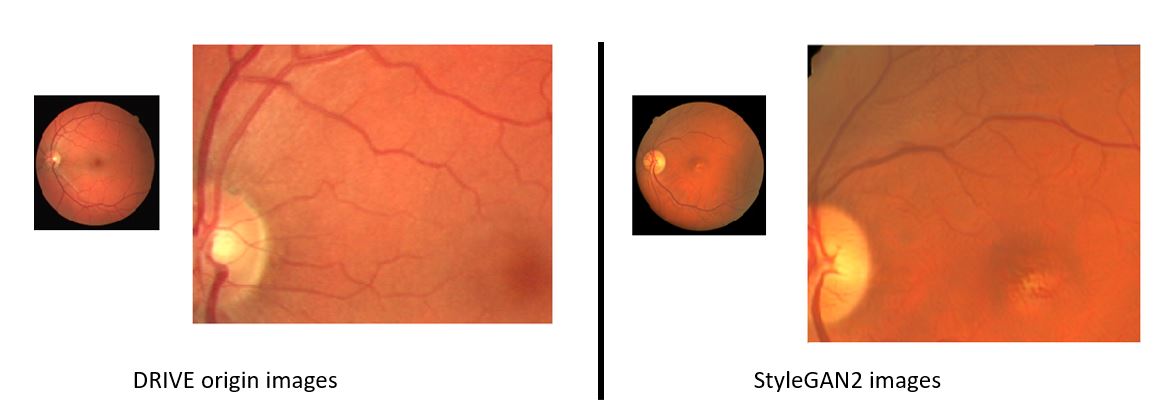}
\caption{\label{fig:Compare2Images.jpg}Compare original image(left) to StyleGAN2 image(right) with zoom in}
\end{figure}

\section{Experiments}

\subsection {Evaluation Metrics}
In order to evaluate the model successful, we compare the segmentation result to the ground truth data, and for each pixel we examine if it True Positive(TP), False Positive (FP), False Negative (FN), or True Negative (TN). Then we define the Sensitivity(SE) which indicates the proportion of the true positive samples among all the predicted positive samples (TP/TP+FP) also known as  Positive Predictive Value (PPV). Specificity (SP) is a metric that measures the proportion of positives that are correctly identified (TP/TP+FN) also known as True Positive Rate (TPR). ACC is a metric for measuring the ratio between the correctly classified pixels and the total pixels in the dataset (TP+TN / TP+TN+FP+FN). Finally there is a unique metric to evaluate segmentation which called F-measure (F1) \cite{sasaki2007truth} that defined by:
\begin{equation}
    F_1 = 2* (PR*SE)/(PR+SE)
\end{equation}
that metric compare the similarity and diversity of the testing datasets.
And for the last - the most important metric is the Area Under the ROC Curve (AUC).

\subsection {Training Details}
The training of both models (StylGAN2 , SA-Unet) have been done using the google colab, pro+ edition.
The preprocessing training include resize of the images so they will be in rectangular size in a power of 2 size, so the size that have been choose is 512*512 pixels.\\
Training StyleGAN2 have been done using ADATorch patch \footnote{https://github.com/NVlabs/stylegan2-ada-pytorch}, using rotate and scale for augmantation.
The metric "fid50k-full" have been chosen for loss metric with the ADAM optimizer for 4K ticks, starting with pre-trained StyleGAN on "ffhq512" dataset. After training we create 1000 images with truncation of 0.7.\\
SA-Unet have been training at start with the parameters of the origin paper \cite{SA-UNet} take 150 epochs, the first 100 epoch have been train with adaptive learning rate start with 0.001, and the last 50 epoch have been trained with learning rate started by 0.0001.
For optimizer we choose ADAM optimizer and the loss chosen to be average of Dice coefficient and binary cross entropy.
The size of the discard blocks of DropBlock is set to 7 as in original paper. \\ 
\\
After training the StyleGAN2 and the SA-Unet we run 100 epochs of SA-Unet from scratch with [50, 200, 500, 1000] synthesised images, learning rate was 0.001, data memory limited us to batch size of 2. After that we done fine tuning with 50 epochs of the original DRIVE data set images. we got 0.98 Accuracy already using the synthesised data only, with 0.96 Accuracy and F1 score of 0.89.\\
After fine tuning we achieved better result, we compare our result to the SA-Unet original result, and present them in table 2.
\\
Next we try all possible combination of training. First the real and then the synthetics (and vice versa), with and without augmentation (only on real, only on synthetics and both) and different number of synthetics images (50, 200, 500 or 1000).

\subsection {Result}

We achieved some interesting working points, each have new SOTA on some metrics but also degradation on another (sometimes very small degradation).
The best configuration of SA-UNET training was to start with the real DRIVE images for 100 epochs with augmentations (color and affine), and then another 50 epochs with 1000 synthetics images without augmentations. The AUC metric we got was 0.977 instead of 0.987 in the original paper, in the ACC we got the same result of 0.968, but the F1 score were to our favor 0.9363 instead of 0.8820 in the SA-Unet original paper. The training process exhibited in figure 5.\\
Behind that in the Precision we got 0.887 which is high above the 0.802 in the original SA-Unet, but in the Sensitivity we decrease the performance from 0.853 to 0.7295 in our model.
When we look for the best F1 score we got 0.983 F1 score, with 0.94 AUC, 0.955 ACC and 0.969 precision.
Another interesting working point we achieved is 0.982 AUC with 0.92 on the F1, 0.9674 in ACC and 0.8726 precision.

After trying all the above configurations (described at 5.2 section) we tested all the networks again but this time together with the original SA-UNET model, in an ensemble model as described above. Due to time limitations we tried only very simple ensemble technique that combine the predictions from both model to one segmentation map. This ensemble give us little improvement in the result. All the result are in table 2. \\

\begin{figure}
\centering
\includegraphics[width=\linewidth]{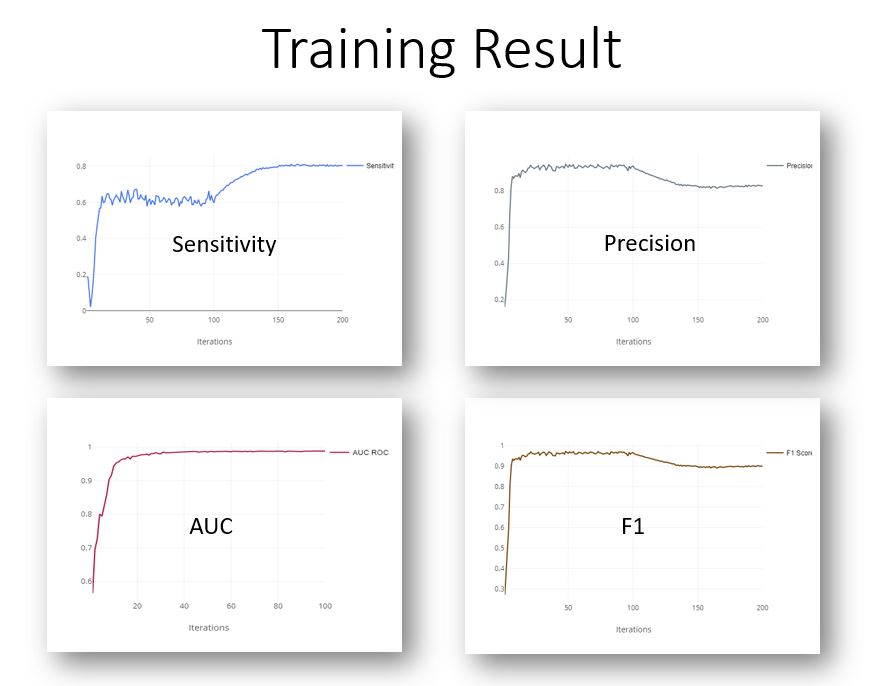}
\caption{\label{fig:Training.jpg}Relevant metrics during train}

\end{figure}

\begin{table}
\centering
\begin{tabular}{p{3.5cm}|p{1.5cm}|p{1.5cm}|p{1.5cm}|p{1cm}|p{1.5cm}|p{1.5cm}}
Method & Sensativity & Specicity & Accuracy & AUC & F1 Score & Precision  \\\hline
SA-UNet &  0.853  & 0.980  & 0.969 & 0.987 & 0.882 & 0.802\\
Only Synthesised data & 0.862 & 0.977 & 0.967 & 0.969 & 0.870 & 0.792\\
First Result & 0.767&0.991&0.965&0.978&0.938&0.889\\
Best F1&0.504&0.998&0.955&0.940&0.983&0.968\\
Ensemble & 0.729 & 0.991 & 0.968 & 0.983 & 0.936  & 0.734\\
Ensample- Sensativity&0.888&0.982&0.968&0.985&0.835&0.734
\end{tabular}
\caption{\label{tab:widgets1}Final performance comparison on the DRIVE dataset.}
\end{table}

\section{Conclusion}

Adding synthetics generated images with GAN can help to get better results when we have lack of ground truth data. Its amazing that train the SA-Unet only on synthetics data can achieve comparable result to the SOTA model that have been trained on real data.\\
With help of StyleGAN, and without any need to manual work on each image, we managed to get SOTA on some metrics while we stay almost with the same result on the remained metrics.
At the same time, we can't ignore that images created by StyleGAN did not manage to reach a good enough resolution in the small details, what cause to loose potential better results. That because the original SA-Unet weakness was in the very small vessels.\\

For future work we want to design more sophisticated ensemble model that can better utilize the advantages of each model - the original SA-Unet and one that trained on synthetics images.\\

\bibliographystyle{alpha}

\bibliography{main}

\appendix{Code:}

The code for the SA-Unet can be found on GitHub: \url{https://github.com/nadavpo/DL_course_project}.\\
This repository include images generated with StyleGan2 (we use StyleGan2 straight forword so thise code not included in our repository) and one can use the train.py or inferance.py after config the parameters on the bottom of each file

\end{document}